\documentclass{Tran-l}
\usepackage{graphicx}
\usepackage[ansinew]{inputenc}
\usepackage[active,new,noold,marker]{xrcs}
\usepackage{lscape}
\usepackage{color}
\usepackage{dcolumn}
\usepackage{bm}
\usepackage{mathrsfs}
\usepackage{dsfont}
\usepackage[colorlinks]{hyperref}
\hypersetup{colorlinks=true,%
            citecolor=blue,%
            filecolor=black,%
            linkcolor=blue,%
            urlcolor=blue}
\def\Note#1{\raisebox{-0.5ex}{$^{\scriptsize\textcolor{green}
{\framebox[0.9\width][c]{\textcolor{blue}{$\,#1\,$}}}\ref{n#1}}$}}
\begin{document}
\ifx\href\undefined\else\hypersetup{linktocpage=true}\fi
\bibliographystyle{apsrev}

\title{The collaboration between Korteweg and de Vries ---
An enquiry into personalities $\phantom{xx}$}

\author{Bastiaan Willink}
\address{Dr Bastiaan Willink \vspace{-6pt}}
\address{Laan van Meerdervoort 1068, 2564 AX, The Hague, The Netherlands}
\email{willinkb@xs4all.nl}

\date{\today}

\begin{abstract}
{\bf In the course of the years the names of Korteweg and de Vries
have come to be closely associated. The equation which is named
after them plays a fundamental role in the theory of non-linear
partial differential equations. What are the origins of the doctoral
dissertation of De Vries and of the Korteweg-de Vries paper?
Bastiaan Willink, a distant relative of both of these
mathematicians, has sought to answer these questions. This article
is based on a lecture delivered by the author at the symposium
dedicated to Korteweg and de Vries at University of Amsterdam in
September 2003.}
\end{abstract}

\maketitle

Since its rediscovery by Zabusky and Kruskal in 1965, an extensive literature has come to exist on the Korteweg-de Vries equation (KdV equation) which describes the behavior of long-wavelength waves in shallow water. It is not the aim of this paper to add to the discussions concerning the contents of this equation or regarding the genesis of the theory of non-linear partial differential equations. Hereto Eduard de Jager has recently added two papers \cite{r1}  Earlier Robert Pego and others questioned the originality of the work by De Vries and Korteweg, especially in relation to the work by Boussinesq \cite{r2}. De Jager has however made it plausible that although the KdV equation can be deduced from an equation of Boussinesq [Joseph Valentin Boussinesq (1842-1929)] by means of a relatively simple substitution, nonetheless Korteweg and De Vries arrived at new and important results through treading a different path than Boussinesq.

In addition to the mathematical and hydrodynamical aspects of the discussion regarding the priority and originality, there are, however, other historical aspects to be considered. Anne Kox has earlier described Korteweg as the nexus between the physics and mathematics departments of University of Amsterdam and Ad Maas has considered him as a transitional figure in the mathematical and academic traditions \cite{r3}. In this paper I shall go into the personal backgrounds of both Korteweg and De Vries which will shed new light on the peculiar genesis of De Vries' doctoral dissertation (\emph{proefschrift}) and the KdV paper. Whereas Korteweg is known as a Dutch pioneer in the area of scientific bibliography, it appears, paradoxically, that some matters may have gone amiss with the digestion of the international literature by his doctoral student (\emph{promovendus}) Gustav de Vries. It appears grotesque that while Korteweg, who has played an important role in the professionalization of the exact sciences in the Netherlands, would have somehow neglected appropriately to supervise one his doctoral students, one of the essential parts of his responsibilities as a university professor. That Korteweg and De Vries were of flesh and blood will become apparent from their biographies to be presented below, which make the remarkable events surrounding De Vries' doctoral research (\emph{promotie}) more understandable.

In a footnote in my book \emph{`De Tweede Gouden Eeuw'} (\emph{The Second Golden Age}) I published a letter from the supervisor [Korteweg] to the doctoral student [De Vries] which at the very least elicits astonishment \cite{r4}. The letter dates from October 1893, one year before the graduation (\emph{promotie}) of De Vries on 1 December 1894. Because of its historical interest, I again cite it here:
\begin{itemize}\sl
\item[{}] ``Dear Sir,\\
To my regret I am unable to accept your dissertation in its present form. It contains too much translated material, where you follow Rayleigh and McCowan to the letter. The remarks and
clarifications that you introduce now and then, do not compensate for this shortcoming. The study of the literature concerning your subject-matter must serve solely as a means for arriving at a more
independent treatment [`whereby you' is put in by mistake, B.W.], expressed in your own words and in accordance with your own line of reasoning, prompted, possibly, by the literature, which should not be followed so literally. When you have mastered your subject-matter to the extent that you can do this, then naturally you will also be confronted with the questions raised by Rayleigh and McCowan, which will provide you with the opportunity to display your strength. \\
In order to facilitate your progress, I send you the outline of a treatment of a single wave according to a slightly modified method due to Rayleigh [according to the submission letter, this very point is central to the final version of the KdV paper, B.W.] [K. would see whether he] could find a guiding principle to offer you for further elaboration [\dots] Naturally, I cannot know whether I can succeed in this. [\dots] For an historical overview of the theory of waves, you should consult much more literature than you have done thus far, and this task will be difficult to carry out in Alkmaar. Your introduction contains too exclusively issues that one can equally well find in handbooks (Lamb and Basset).\\
It is obviously a disappointment for you who must have deemed to have already almost completed your task, to discover that you have apparently only completed the preparatory work. In the meantime do not be down-hearted. With pleasure I will do my best to help you mount the horse [\dots]''\Note{2}
\end{itemize}

From a study by Eduard de Jager of the preserved records of De Vries' it would appear that the latter subsequently almost independently arrived at the results presented in his doctoral dissertation.\Note{3} It appears therefore that De Vries deserves more credit than I have given him in 1998. It remains however striking for a thesis advisor (\emph{promotor}) criticising a doctoral student one year before graduation ceremony (\emph{promotieplechtigheid}) on account of the latter having advanced too little of his own ideas. At the very least De Vries must have been under considerable pressure in the period after the letter. The circumstances appeared so peculiar to me, that after the publication of my book I did some fresh investigations. The De Vries archive that was discovered as a result, has subsequently enabled Eduard de Jager to appraise the contributions of De Vries better. I myself have discovered other facts and documents which make it more comprehensible, why precisely around 1894 Korteweg and De Vries were pressed for time, whereby in some respects they acted very rashly. \\

What has happened after Korteweg's emphatic letter? And is it possible to obtain indirect evidence concerning De Vries' talent, so that it can be made more comprehensible whether he nevertheless did original work in the last year of his doctoral research? Has he achieved other things? And if this was not the case, how was it possible that Korteweg was able to lead him so far that De Vries surpassed himself? And finally, it was important to unearth what Korteweg and De Vries knew of the closely related work by Boussinesq. Because such facts have often the tendency to disappear without trace, I give details that somewhat colour in the sketches of two diverse personalities while fighting with stress and time.

Delving into archives of Korteweg was straightforward, finding that of de Vries amounted to research of half a year. Concerning Korteweg, many of his documents have been preserved. The most important of all archives is his scientific archive in University Library of University of Amsterdam.\Note{4} What can we learn from this material about Korteweg in general and in particular over the period around 1894? The picture is fairly complete. I shall attempt to present a somewhat detailed impression of Korteweg the man, so that his responses to the complications regarding the doctoral thesis during 1893-1895 can be better appreciated.

\section*{Life of Korteweg}

\begin{figure}[t!]
\begin{center}
\includegraphics[width=2.5in]{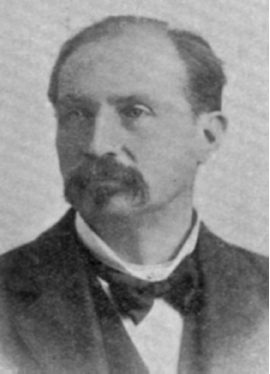}
\caption{\label{f1} Diederik Korteweg (1848-1941) around 1898}
\end{center}
\end{figure}

Diederik [Johannes] Korteweg (1848-1941) was the eldest son in a family of six children, five boys and one girl. His father was a non-Catholic district Judge in Den Bosch, the provincial capital city of the mainly catholic province of Noord-Brabant in the south of the Netherlands. Consequently, the family was in a rather isolated social position. It makes one think of the father of the painter Vincent van Gogh (1853-1890), who was a protestant vicar in Nuenen. Perhaps this position contributed to the father's decision to become a freemason, a move that was not unusual amongst progressive citizens of this period.\Note{5} In this milieu he made his real career. He made it to a member of the Dutch Central Committee (\emph{het Nederlands hoofdbestuur}) under Prince Frederik (1797-1881), the brother of  King Willem II, as well as of international arbitrage commissions.\Note{6} He succeeded in passing on his independent and international outlook to his sons. I have not been able to find anything concerning membership in freemasonry of any of his sons. They were all unbelievers and politically of the left-liberal brand. Possibly as scientists they were not greatly charmed by the semi-religious aspects of freemasonry. It is known that Died[erik]  played a role in the Amsterdam liberalism and that he financially helped the well-known Dutch novelist Multatuli (Eduard Douwes Dekker, 1820-1887). The second son, Bas[tiaan], came also in direct contact with Multatuli. He was married to Elize Baart who had played in the premi\`ere of Multatuli's play \emph{Vorstenschool} (School for Monarchs). In politics he went farther than Died and became a socialist. When he began to despair of future, he together with his wife committed suicide in 1879, a tragedy of which the news reached the national press but especially deeply affected the family. The writer Jeroen Brouwers has devoted a small book to this event.\Note{7}

Died's third brother, Jo[han], became a professor of surgery and his fourth brother, Piet[er] a notable malaria research scientist.\Note{8} I am of the opinion, but Brouwers considered it too speculative, that the strongly competitive atmosphere prevailing amongst the brothers may have partly contributed to the suicide of Bas. As students they gave each other puzzles from their own fields, so as to compel each other to study other subjects in addition to their own.\Note{9} Bas was also a mathematician and owed his position as a lecturer at The Royal Military Academy (\emph{Koninklijke Militaire Academie}) in Breda to Died, but was not able to accomplish his own independent research. Later we shall see that a similar situation prevailed in De Vries' equally respectable-bourgeois family.

Died Korteweg was therefore agnostic, liberal and had grown up in a family where members were expected to achieve. It is slightly peculiar that he has had a relatively precarious education. He received his elementary education first at Institute Berman (\emph{Instituut Berman}) in Den Bosch and later took private lessons from the school inspector Ringeling. Only in September 1865 did he go to Polytechnical School (\emph{Polytechnische School}, later to become Delft University of Technology), which according to the regulations of 1863 was classified as an institute of Secondary Education (\emph{Middelbaar Onderwijs}, MO). He came to experience the curriculum in Delft as too applied and unscientific. After his Secondary-School teachers examination (\emph{MO examen}), in 1869, at age twenty-one, he became teacher at a High School (\emph{Hogereburgerschool, HBS}, later and until 1968 a preparatory school for academic education without Classical languages) in Tilburg and subsequently in Breda; he continued studying mathematics in his leisure time. Only at his twenty-eighth, in 1876, did he take the university entrance examination in Utrecht. When in April 1877 Died passed his Bachelor of Science (\emph{kandidaats}) examination, it was one month before his three-year younger brother Jo obtained his PhD (\emph{promoveerde}) in medicine.\Note{10} However, already since 1877 Diederik Korteweg assisted the physicist, and the future physics Nobel laureate [1910], Johannes Diederik van der Waals (1837-1923) with solving mathematical problems. Shortly afterwards, on 31 January 1878, he passed with honours (\emph{cum laude}) his Master of Science (\emph{doctoraal}) examination in Amsterdam. By this time he must have advanced far with his doctoral research, concerning speed of propagation of waves in elastic tubes, since he obtained his PhD (\emph{promoveerde}) in the same year, on 12 July, as the first doctoral graduate of University of Amsterdam. According to his own statement, he had received much assistance from Professor Van den Berg from Leiden.\Note{11} Two weeks after his doctoral examination he married with a Baroness d'Aulnis de Bourouill, at first sight an instance of `marrying upwards', but the d'Aulnises were not very class-conscious, and the brother of Bientje d'Aulnis was the pioneer of mathematical economy in the Netherlands. To their disappointment, the pair remained childless.

We now look yet again into the career of Diederik Korteweg. Astonishing, such a slow and then suddenly swift career: in 1876 entrance examination, in 1877 his BS examination and immediately afterwards assistant to Van der Waals, in 1878 his MS and PhD examinations. In 1881 Korteweg was appointed a university professor.\Note{12} The issue of `competing brothers' played, in addition to the aspiration to become a university professor, a role in the strong drive towards accomplishment, as it was also the case with De Vries and later with Brouwer (Luitzen Egbertus Jan Brouwer, 1881-1966), the latter competing specifically with his cousins. Following Korteweg's appointment in 1881, he gave lectures in mathematics, mechanics and astronomy. This he did meticulously; he could be very critical of students, but later he was also certainly a fatherly figure. In addition to teaching, he worked, encouraged by Van der Waals, on mathematical description of plaits on surfaces and the application of this to Van der Waals' theory concerning equilibrium phases of binary mixtures. The extensive papers of Korteweg on this subject appeared in 1889 and 1891, and have been rediscovered recently by Mrs Sengers-Levelt and put in their historical context \cite{r5}. Alongside the work by P.~H. Schoute (Pieter Hendrik Schoute, 1846-1923), these were the first mathematical papers of international significance in the Second Dutch Golden Age. Ad Maas has rightly pointed out that in some respects Korteweg can be considered as a transitional figure between professors who around the middle of the nineteenth century did useful applied mathematics, and the genuine researchers, such as Brouwer, after 1900.\Note{13} Nonetheless, in his role as a mathematical researcher Korteweg appears to be more significant than most `professionals' of the generation succeeding his.

What is striking about Korteweg is the considerable amount of energy with which he has worked in many diverse areas, and that he almost always has obtained results. He did most of the editorial work for the collected works of Christiaan Huygens (1629-1695) from 1911 to 1927, during which time he made discoveries concerning the influence of Snellius (Willebrord Snellius, 1580-1626) on Descartes (Ren\'e Descartes, 1596-1650) and of the latter on Huygens. In the meantime he simultaneously plunged into work in the regional Dutch office of International Catalogue of Scientific Literature (\emph{Internationale Catalogus van wetenschappelijke literatuur}), where all scientific papers and books were registered and classified. In the doctoral thesis of Paul Schneiders dealing with the library and documentation movement in 1880-1914, Korteweg's Sisyphean task is discussed in detail \cite{r6}.

\section*{The doctoral graduation of de Vries}

The first directly relevant event for Korteweg in the period of stress surrounding the graduation of De Vries, was the death of the city archivist of Amsterdam, Nicolaas de Roever (1850-1893). Because De Roever's spouse had already died, their surviving three children, two girls and a boy, were adopted, in 1893 or 1894, by Korteweg and his wife who were childless and in their forties. This must have created quite some complications. It must also have been a blow to the family that the eleven-years-old adoptee-son Arend died in 1896; I am not aware whether or not Arend had been ill long before his untimely death.\Note{14}

A second directly relevant circumstance preceding the graduation of de Vries, besides his sudden paternity, was that Korteweg in 1893-1894 was Vice-Chancellor (\emph{rector magnificus}) of University of Amsterdam and amongst other things worked on the text of his oration \emph{The Golden Age of mathematics in the Netherlands} (`\emph{Het bloeitijdperk der wiskundige wetenschappen in Nederland}'), in which he anticipated his historical research during the second half of his scientific life. For completeness, with `the golden age' (`het bloeitijdperk') he referred to the seventeenth century. He delivered his Vice-Chancellor's oration in January 1894. All these adopter's, vice-chancellor's and professor's (he continued to carry out his teaching duties) time-consuming responsibilities clarify why he was rather blunt when he received the first draft of De Vries' thesis. The doctoral advisor Korteweg was an arduous, sober and efficient worker of great versatility who nonetheless had rather reached the limits of what he was capable of. He looked back on his own efficient doctoral research and saw that De Vries did not really make headway. In the past years he had dealt with the history of the Dutch mathematicians of an earlier age and the theory of analytic surfaces, and presumably had not closely followed the literature on hydrodynamics. Therefore, while he saw well that De Vries had studied Scott Russell (John Scott Russell, 1808-1882), Airy (George Biddell Airy, 1801-1892), Rayleigh (John William Strutt, 3rd Baron Rayleigh, 1842-1919), McCowan (John McCowan, 1863-1900), Greenhill (Sir Alfred George Greenhill, 1847-1927) and Boussinesq, he did not see that important publications by the latter author were missing.

A comparable situation arose ten years later with Brouwer. The latter also submitted a draft thesis with which Korteweg had much trouble. What a heap of unnecessary twaddle! Again, in 1905, Korteweg became angry about the excessive amount of inefficiency and compelled Brouwer to remove all kinds of philosophical passages from his thesis. The difference with the case of De Vries was that Brouwer's most important ideas were already contained in his draft thesis, while that was certainly not the case with De Vries.

\section*{The life of de Vries}

\begin{figure}[t!]
\begin{center}
\includegraphics[width=2.5in]{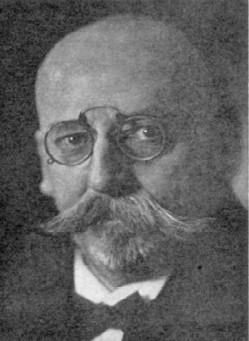}
\caption{\label{f2} Gustav de Vries (1866-1934)}
\end{center}
\end{figure}

How was the situation on the other side of the graduation ceremony? What do we know about the doctoral student De Vries? A book from 1936 about the De Vries-family was a good starting point for further research.\Note{15}\Note{16} It turned out that the Utrecht university professor of mathematics Jan de Vries was Gustav's elder brother.\Note{17} A third brother, August, became Secretary-General of the Ministry of Finance and Privy Councillor Extraordinary (\emph{Staatsraad in buitengewone dienst}).\Note{18} Two brothers who went nearly as far as two of the Korteweg brothers. The branches of the family tree corresponding to these successful brothers, who must have been cause for some difficult to measure frustrations for Gustav, have died out; this is certainly not the case with Gustav's family branch. It turned out that even two grandsons of Gustav's live in Leiden, who are in possession of an archive of their grandfather. What kind of picture does one obtain from this archival material and other documents?

Gustav de Vries came from a family that in many respects is reminiscent of that of Korteweg, although a father who was a bookseller in Amsterdam considerably differs from a judge in Den Bosch. Gustav's elder brother was, as said earlier, a mathematics professor, just as the eldest son of the Kortewegs. Died Korteweg and Jan de Vries knew each other well, they were both members of the Dutch Royal Academy of Sciences (\emph{Koninklijke Akademie der Wetenschappen}). And both of them had a younger brother talented in mathematics whom each helped in obtaining a position. In the case of Korteweg, this was at the Royal Military Academy (\emph{Koninklijke Militaire Academie}, KMA) in Breda, and in the case of Jan de Vries, a position at the five-year HBS school in Haarlem, which he was able to pass down to Gustav when he himself went to Delft Polytechnical School. But Gustav de Vries reminds one of Bastiaan Korteweg also in other respects. He was not able to find his way in this position.

After his graduation he published with difficulty and that meant that he stuck to this HBS position. The first years went well, and he even stood in for an ailing colleague. But in the 1902, 1903 and 1904 annual reports of the HBS one repeatedly encounters passages like ``{\sl absent for a considerable period because of illness}''.\Note{19} In an open letter from 1908 --- also Bas Korteweg published a pamhlet about his resignation from the KMA --- of de Vries to the Alderman of Education (\emph{wethouder van onderwijs}) in Haarlem, Thiel,\Note{20} he explains that in 1902 he had been for five weeks in a sanatorium for patients suffering from nervous breakdown (\emph{zenuwlijders}) as a result of disappointments concerning failed job applications and for lack of support in this from his principal Brongersma.

\section*{The fight for recognition}

De Vries did try to continue studying and writing in his private library at Ripperdapark 45, where the four children, who followed a deceased first child, were not particularly welcome. In this way he published two papers in 1900 concerning cyclones in the \emph{Proceedings of the Royal Netherlands Academy}, which I have not looked into, and in 1907 a \emph{Concise textbook on arithmetic and algebra} (\emph{Beknopt leerboek der reken- en stelkunde}) published by \emph{De Erven Bohn}. In 1909 he was discharged from the five-year HBS school and appointed as a teacher at the three-year variant of this school. The year before, he must have been unpleasantly surprised when he was returned the manuscript that he had submitted through Korteweg to \emph{Nieuw Archief voor Wiskunde} (\emph{New Archive for Mathematics}). The Editor, the university professor and algebraist Kluyver from Leiden, writes to Korteweg, who showed the letter to De Vries, that:
\begin{itemize}\sl
\item[{}] ``Amice [Dear Colleague, the word is derived from the Latin word `amicus', B.W.], \\
    I have perused the strange piece by mister De Vries. It gives me the impression that the author has accidentally noticed a quite natural and unexceptional phenomenon, of whose true nature he makes no correct representation, and now more or less raises the status of what actually amounts to a commonplace thing to a miracle.''
\end{itemize}
Subsequently, he gives an explanation of the weak aspects in the manuscript and calls De Vries a ``pretty old author'' (``\emph{schrijver op leeftijd}'') (de Vries was then $42$!).\Note{21}

The letter of Kluyver is from April 1908, De Vries' heated open letter to the Alderman from August of the same year. In this letter he writes about his frustrations and lack of support from his superior, and interestingly states: ``[this] added to the grief, caused by the sudden death of a child''. The disappointments came in the first place, which says something about the immensity of his frustration. From the documents it appears that De Vries perhaps made too high demands on his pupils in the higher classes, but that he did distinctively well in the lower ones. This signifies that he as yet did not expect much from the younger students, but that he perhaps looked for spiritual affinity with the older ones. At the time when he was required by his director to teach below his level of qualification, he applied to all kinds of positions, but did not even get nominated. All exceedingly frustrating. Also, from the long, even printed, open letter to the Alderman it appears that there were many colleagues who found him socially incompetent in dealing with punishing pupils, but also in regard to teaching itself. The picture is that of a man with considerable communication problems.\Note{22}

In spite of all setbacks, De Vries published in 1912, through Korteweg, two papers about his own `\emph{calculus rationis}' in \emph{Proceedings of the Royal Netherlands Academy of Sciences}.\Note{23} After 1912 he had new life-fulfilling experiences. He felt suitably at home in the new school where little demands were made on him. He taught for seven hours mathematics in all the three classes and for this purpose he used his own textbook until 1911 (not after this date) and also that of his brother and Janssen van Raaij about planar geometry. In addition, he gave lessons in accounting. In 1913 he was confirmed as a member of the freemason Lodge \emph{Vicit vim virtus} (Virtue has overcome the power), of which he became the Master in 1916.\Note{24} Together with some others, in 1916 he seceded from this Lodge and joined the new Lodge \emph{Kennemerland}. He held for his Lodges long and elaborate discussions on philosophical subjects. A remarkable text in his inherited papers concerns a long analysis of Goethe's \emph{Faust}, in which he comports with a French commentator.\Note{25}

In the end De Vries became a spiritualist. This may ring somewhat strange nowadays, however around 1900 there were many scientists, especially in Britain and America, who did intensive research concerning psychics and theorized about the fourth and higher dimensions so as to be able to explain the possibility of life after death, such as the French Nobel laureate in medicine Richet (Charles Robert Richet, 1850-1935) and the physicists Oliver Lodge (Oliver Joseph Lodge, 1851-1940), who proposed the term `black hole', and William Crookes (Sir William Crookes, 1832-1919), of whom De Vries had studied papers in the course of his doctoral research. In December 1934, following a s\'eance in Haarlem-Noord, De Vries was knocked down by a car;\Note{26} he died later in hospital as a consequence. His ever sickly wife survived him for three years. It seems likely that by now almost everything about De Vries and his sorrowful career is unearthed.\Note{27} He was a decently good researcher, who however with his doctoral dissertation straightaway produced his most significant achievement. Either directly or indirectly, this must have been owing to the pressure put on him by Korteweg and to his experiencing his teaching obligations as onerous.

\section*{Stress and carelessness}

\begin{figure}[t!]
\begin{center}
\includegraphics[width=4.2in]{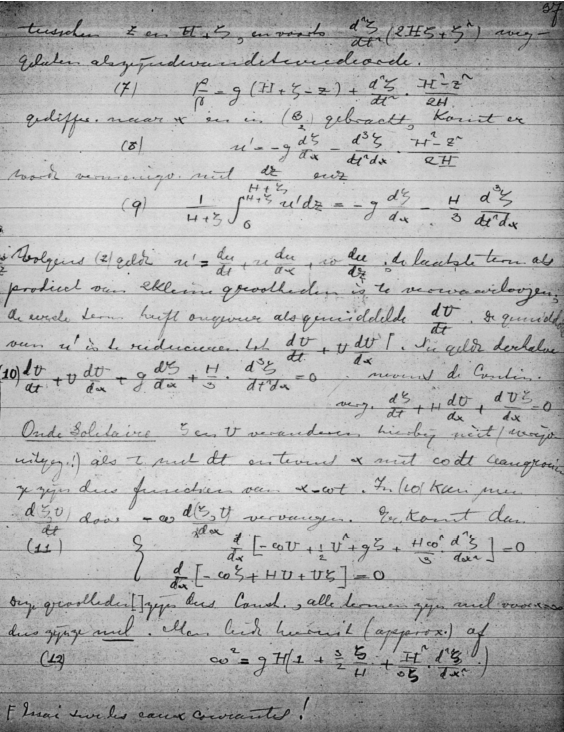}
\caption{\label{f3} Mentioning by De Vries of Boussinesq's `\emph{Essai}' [treatise] at the bottom of page $37$ of his summary of Saint-Venant, 1885.}
\end{center}
\end{figure}

What were De Vries' circumstances in 1893-1894, what kind of interaction took place between him and Korteweg, and what about their knowledge of the work by Boussinesq? We have seen that De Vries, just before his elder brother Jan arranged for the new position in Haarlem, had submitted the first draft of his doctoral thesis to Korteweg. This was followed by Korteweg's rebuke that made De Vries shift into a higher gear. In the meantime the HBS-teacher was already too busy. After his graduation, in the time that the paper for the Royal Society was being prepared by ``my young friend and myself'' (writes Korteweg in his letter of submission to the Royal Society),\Note{28} De Vries complains that he no longer has spare time for either writing or doing research, as a result of the large amount of tests that he had to mark (but this must have already been the case earlier), while Korteweg in his letter had just insisted on much more scholarship. Thesis advisor \emph{and} research student were therefore under considerable pressure at the beginning of 1894, had both little time and must have thought: it is now or never.

\begin{figure}[t!]
\begin{center}
\includegraphics[width=3.2in]{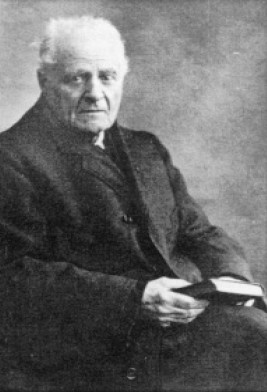}
\caption{\label{f4} Joseph Valentin Boussinesq (1842-1929)}
\end{center}
\end{figure}

This appears to have affected the study of literature. Unfortunately, De Vries has not included any bibliography in his doctoral thesis. He has however left behind some neat excerpts from the literature studied by him, amongst which papers of Boussinesq from 1870 and 1871.\Note{29} We can closely follow him in his readings. He began with the English literature about his subject of interest and must have encountered Boussinesq's name in Rayleigh's paper \emph{On Waves} from April 1876 in \emph{Philosophical Magazine}. Rayleigh writes:
\begin{itemize}\sl
\item[{}] ``I have lately seen a memoir by Mr. Boussinesq (1871, {\it Comptes Rendus}, Vol. LXXII.), in which is contained a theory of the solitary wave very similar to that of this paper. So far as our results are common, the credit of the priority belongs of course to Mr. Boussinesq.''
\end{itemize}
From the phrasing it appears that Rayleigh did not know that Boussinesq was already a university professor, or that the third Baron did not find this notable.

De Vries subsequently also studied some French publications, Boussinesq 1870 and the first paper from 1871, but also papers by Saint-Venant (Adh\'emar Jean Claude Barr\'e de Saint-Venant, 1797-1886) from 1885. In one of Saint-Venant's papers studied by De Vries, published in \emph{Comptes Rendus}, Vol. CI, from 1885, one finds a clear reference to Boussinesq's 680-pages treatise from 1877, \emph{Essai sur la th\'eorie des eaux courantes}, wherein, as Pego and others have pointed out, the KdV equation occurs in a footnote. De Vries subsequently explicitly refers to Boussinesq's \emph{Essai} on pages 38 and 40 of his excerpt of Saint-Venant's paper and even writes at the bottom of page 37 ``{\sl Essai sur les eaux courantes}'', followed by an exclamation mark! (See Fig.~\ref{f3}.)\Note{30}

In his extensive paper, De Jager posits \cite{r1}:
\begin{itemize}\sl
\item[{}] ``It is somewhat surprising that Korteweg and de Vries refer in their paper only to Boussinesq's short communication in the {\it Comptes Rendus} of 1871 and not to the extensive article in the {\it J. Math. Pures et Appl.} and the {\it Essai sur la th\'eorie des eaux courantes} of 1872, respectively 1877. However, we should realize that the international exchange of scientific achievements in those days was not at the level as it is today''\Note{31}
\end{itemize}
This seems jumping to conclusions. There are, in spite of Korteweg's exhortations to study more, no indications that De Vries has been aware of Boussinesq's paper of 1872. There is also nothing that would suggest that he has read the treatise from 1877, which is odd, especially if you put an exclamation mark in front of its title. Very possibly there was at that time already a copy of the work (which received a Prize from \emph{Acad\'emie des Sciences} \cite{r7}) in the Netherlands; for example, I found the treatise in the library of Polytechnical School in Delft, where it must have arrived not long after its publication in 1877. It is also possible that he strayed off in the book --- it is a voluminous and difficult book --- but then De Vries should have discussed it with Korteweg, and of this there is no trace to be found in the dissertation, the correspondence and the De Vries-archives. Finally, it is strange that Korteweg should not have been informed about the crowned treatise through a different channel.

\section*{Comedy of errors}

In any case, being under pressure, De Vries has, consciously or unconsciously, ignored something of significance. Notwithstanding Korteweg's considerable attention to bibliographically responsible research, which is amply evident from his letter quoted above, in which additional literature study was explicitly asked for, and De Vries' carefully-prepared excerpts, it appears that no study of Boussinesq's works of 1872 and 1877 followed the exclamation mark. In the case of Rayleigh, it was already to some extent peculiar that by 1876 he had not already investigated the literature of 1872. But then Boussinesq's major work from 1877 and Saint-Venant's citations from 1885 had not as yet been published. The omission by De Vries in 1894 seems to me unforgivable, even though at the time he and Korteweg were possibly following a different line of theoretical enquiry.

It is equally strange that Boussinesq, who lived until 1929, has never made any complaints to Korteweg or the editor of \emph{Philosophical Magazine}. Saint-Venant did not experience this `comedy of errors'. In 1885 he was no less than 88 years old and died the following year. Earlier, he had played a significant role in the life of Boussinesq and helped him with his appointment to his first professorial position in Lille, where also Pasteur (Louis Pasteur, 1822-1895) had begun his career. Boussinesq had looked forward to collaboration with the energetic old man, when he in 1886 moved to Sorbonne, but that unfortunately was not to be. Later Boussinesq became Dean of the natural sciences department of Academy of Sciences (\emph{Acad\'emie des Sciences}). He certainly had an international network. It should be investigated whether there has existed a direct or indirect contact between Boussinesq and Korteweg.\Note{32} The latter has for instance exchanged some correspondence with Paul Appell (Paul \'Emile Appell, 1855-1930), who must have known Boussinesq well.

While Korteweg and De Vries were evidently internationally oriented, for the neglect by Rayleigh of the 1872 paper of Boussinesq and the absence of response by the latter on the KdV paper from 1896, we should take into account that at the time in large countries the national scientific cultures were dominant. Rayleigh read particularly the works by the British and the editors of \emph{Philosophical Magazine} did not know the international literature sufficiently well. In his turn, Boussinesq read above all works by the French. Occasionally they looked over the national borders, and then to Germany. There the first two volumes of the \emph{Poggendorff} bibliography, \emph{Wörterbuch} (Johann Christian Poggendorf, 1796-1877), which were above all of historical nature, had
already been published. Also, \emph{Catalogue of Scientific Papers} of the Royal Society did exist already, but was evidently not consulted by every professional researcher.

It is certain that Korteweg and De Vries, living in a relatively small country, were more internationally oriented than Rayleigh and Boussinesq, and Korteweg was moreover a good bibliographer. However, under the pressure of time they failed to function properly. De Vries ceased to study Boussinesq subsequent to having flagged an exclamation mark in his notes indicating the significance of Boussinesq's work (see Fig.~\ref{f3}). It is conceivable that he subsequently did not inform Korteweg about the existence of the extensive treatise, even if Korteweg would have already considered that all the rewriting had been more than enough. Apparently, also he, even despite his insistence by the end of 1893 on much more study of literature, made a mistake through the bustle. The pressure exerted by Korteweg and the circumstances made that De Vries surpassed himself, which is the positive side of the affair. Of course, we should be glad with the doctoral dissertation of De Vries and the KdV paper, but it is certainly conceivable that De Vries would not have graduated after having deeply studied Boussinesq's pertinent publications'. $\hfill\Box$

\section*{Notes}
{\small

\noindent
\subsection*{\label{n1} 1}
I thank Mrs Sengers-Levelt for commenting on an earlier draft of this paper, which is based on a lecture, delivered by me in September 2003 at the symposium dedicated to Korteweg and De Vries at University of Amsterdam (UvA), and Dr Behnam Farid for his insightful comments concerning De Vries' reference to Boussenesq's \emph{Essai} in the footnote on page 37 of De Vries' excerpt, shown in Fig.~\ref{f3}.

\vspace{-0.2cm}
\noindent
\subsection*{\label{n2} 2}
\emph{The Korteweg Archives at University of Amsterdam} is primarily divided into two main parts, the Huygens Edition and the rest, and secondarily ordered chronologically. I have not been able to find Korteweg's own outline to which he refers, either in his own or in the De Vries archives in Leiden, maintained by De Vries' grandsons.

\vspace{-0.2cm}
\noindent
\subsection*{\label{n3} 3}
Communicated to me in writing by Eduard de Jager.

\vspace{-0.2cm}
\noindent
\subsection*{\label{n4} 4}
After Korteweg's death, his pupil Gerrit Mannoury arranged Korteweg's correspondence according to the alphabetical order of correspondents' names. This he did in the Korteweg house built by Pierre Cuypers, which includes a two-stories conservatory on the corner of Vondelstraat opposite the \emph{Heilige Hartkerk} (the Holy Heart Church), where for a long time he himself was taught by Korteweg in preparation for university education. Unfortunately, the archivist at Amsterdam University Library has rearranged the entire archive, this time ordered chronologically per year, so that I had to spend many hours searching for the Korteweg-de Vries correspondence in the entire archive. The archive contains also `the Mountain of Brouwer', the many letters from and to Brouwer around 1906 (fortunately Korteweg preserved his own drafts), which Dirk van Dalen has used in his books. Also, Mrs Sengers-Levelt has searched through the archive for her book \cite{r5} concerning thermodynamics in the Netherlands around 1900.

\vspace{-0.2cm}
\noindent
\subsection*{\label{n5} 5}
There exists no scientifically rigorous book on the history of freemasonry in the Netherlands (however, since not long ago there is a Chair at University of Leiden). It is well-known that for example Multatuli and Jacob van Lennep [1802-1868, poet and novelist] were members. Under the patronage of Prince Frederik, the Lodges were only open to men; they were meeting places for in particular members of the upper-classes.

\vspace{-0.2cm}
\noindent
\subsection*{\label{n6} 6}
Private archival records, originating from my grandmother, Mrs W.~M. van Steeden-Korteweg.

\vspace{-0.2cm}
\noindent
\subsection*{\label{n7} 7}
Jeroen Brouwers, \emph{Twee verwoeste levens. De levensloop en de dubbelzelfmoord van Elize Baart en Bastiaan Korteweg} (\emph{Two destroyed lives. The biography and the double-suicide of Elize Baart and Bastiaan Korteweg}), Amsterdam (1993).

\vspace{-0.2cm}
\noindent
\subsection*{\label{n8} 8}
Johan was my great-grandfather. I received a great deal of Korteweg-material through my grandmother and my uncle Jaap van Steeden, her only son. The fourth Korteweg-brother Piet had a son named Remmert, later to become a major pioneer in the area of cancer research. The fifth brother, Willem, died while studying Philosophy at Leiden University. Finally, there was a sister, Mrs Rand-Korteweg, of whom almost nothing is known.

\vspace{-0.2cm}
\noindent
\subsection*{\label{n9} 9}
Notes from my grandmother, Mrs van Steeden-Korteweg.

\vspace{-0.2cm}
\noindent
\subsection*{\label{n10} 10}
Jo also didn't have a gymnasium diploma; he too had received private education.

\vspace{-0.2cm}
\noindent
\subsection*{\label{n11} 11}
\emph{Onze Hoogleraren} (\emph{Our Professors}), Rotterdam, 1898, page 256 (brother Johan follows on page 267).

\vspace{0.2cm}
\noindent
\subsection*{\label{n12} 12}
Brother Jo just in 1887.

\vspace{-0.2cm}
\noindent
\subsection*{\label{n13} 13}
Ad Maas, \emph{op. cit.}

\vspace{-0.2cm}
\noindent
\subsection*{\label{n14} 14}
Korteweg and Bientje d'Aulnis are buried with him in Zorgvlied, a cemetery in Amsterdam

\vspace{-0.2cm}
\noindent
\subsection*{\label{n15} 15}
The search after his antecedents was a much more time-consuming task than was the case with Korteweg. There were more mathematicians with the name De Vries, but of course they were not all relations. Already after a short time it became evident that Gustav de Vries was no relation of Korteweg's junior colleague Hendrik de Vries. The golden hint came from De Vries himself. In his doctoral dissertation he had to mention that he was ``born in Amsterdam''. This, added to his special first name, led in the municipality archives, by way of the ten-year indexes (\emph{tienjarige tafels}), to his parents and his date of birth in 1866. Through the Central Bureau of Genealogy in The Hague, the following relations were found with the aid of person's cards (\emph{persoonskaarten}): following a few weeks of investigations, two elderly women from the Veluwe [a district in the Province Gelderland] appeared to be able to tell a great deal about Gustav de Vries, Mrs de Vries-Helmert from Ermelo, 87 years old in 1999, who had been married to David de Vries; Ms Rie Bosman from Harderwijk, 73 years old in 1999 and partner of the late Rubertus Jan de Vries (Ruurd). These women explained that the family had counted amongst its members both followers of right-wing and anti-Semitic convictions (Gustav appears to have been one of them) and left-wing resistance fighters against the German occupation of the Netherlands during the World War II.

\vspace{-0.2cm}
\noindent
\subsection*{\label{n16} 16}
\emph{Een Friesch geslacht uit Amsterdam} (A Friesian clan from Amsterdam), 1936 (not available in shops but found in libraries).

\vspace{-0.2cm}
\noindent
\subsection*{\label{n17} 17}
The family branch of Jan de Vries died out, when his grandson Jan (born in 1920, as son of Dr Jan de Vries, mathematics teacher at the \emph{Amersfoorts Gymnasium}) together with other 49 resistance fighters were executed by the occupying German forces at Varsseveld [in the \emph{Gelderland} Province] on 2 March 1945 in retaliation for an assassination in the neighbourhood where 4 German military personnel had been killed (Source: The Netherlands Institute for War Documentation, \emph{Nederlandse Instituut voor Oorlogsdocumentatie}).

\vspace{-0.2cm}
\noindent
\subsection*{\label{n18} 18}
When I had a meeting with the grandsons of De Vries at a small canal in Leiden, I sensed some degree of distrust. A family member of Korteweg, according to some family tradition in the De Vries family, turned out to have ran away with ideas of their grandfather! I had just behind me a prolonged self-imposed impossible mission, to check, after completion of my book, once and for all, if I had not done injustice to De Vries, so that I felt very uncomfortable. But some gaps were somewhat later filled in by Gustav's brother August. It `transpired' that August, who bought a double name and was later called August Laman de Vries, through the Roelvink family was an uncle of my grandfather Willink. In this way at least one traceable genealogical connection between Korteweg and de Vries turned out to exist, and I was able to point out that my family loyalty was not all too one-sided.

\vspace{-0.2cm}
\noindent
\subsection*{\label{n19} 19}
The Municipal Archives Haarlem (\emph{Gemeente Archief Haarlem}) (since merged with the \emph{State Archives in North-Holland} (\emph{Rijksarchief in Noord-Holland})), Archives of Predecessors of the \emph{Spaarne} Schools in Haarlem (\emph{Archieven rechtsvoorgangers Spaarne Scholengemeenschap} te Haarlem), Archives of the HBS-B, 1864-1958, inv. nrs. 35 and 36, Annual Reports 1889-1906 and 1907-1920.

\vspace{-0.2cm}
\noindent
\subsection*{\label{n20} 20}
This was contained in the documents held by De Vries' grandsons.

\vspace{-0.2cm}
\noindent
\subsection*{\label{n21} 21}
The letter originates from the Korteweg Archives at University Library of Amsterdam University. Lourens van den Brom, with whom I came in touch through his archival researches pertaining to Korteweg's academic career, has examined the arguments of Kluyver and made them plausible.

\vspace{-0.2cm}
\noindent
\subsection*{\label{n22} 22}
The modern specialists would perhaps speak of the Asperger syndrome.

\vspace{-0.2cm}
\noindent
\subsection*{\label{n23} 23}
These papers have been considered by Professor Anne Troelstra, who similarly found that they did not go deeply into research by others. They have nothing to do with fundamental research. ``{\sl Perhaps someone with analytical expertise can find something more in these. My impression is that Korteweg has encouraged De Vries to build on the works of others in his doctoral research, but that De Vries has subsequently been working in too isolated an environment.} [Translated from Dutch, B.W.]''

\vspace{-0.2cm}
\noindent
\subsection*{\label{n24} 24}
I have not been able to find out why he, already in 1908, used the word ``brother'' in a letter, but it looks as though he was a freemason earlier. The data concerning freemasonry originate from The Cultural Masonic Centre (het \emph{Cultureel Ma\c{c}onniek Centrum}) Prins Frederik, in The Hague.

\vspace{-0.2cm}
\noindent
\subsection*{\label{n25} 25}
This piece is also with the brothers De Vries in Leiden.

\vspace{-0.2cm}
\noindent
\subsection*{\label{n26} 26}
Source: Mrs de Vries-Helmert.

\vspace{-0.2cm}
\noindent
\subsection*{\label{n27} 27}
Troelstra suggested rightly that an analysis expert should be asked to consider the `\emph{calculus rationis}' publications. In addition, it should be certainly worthwhile to make a detailed comparison between De Vries' doctoral dissertation and the KdV paper.

\vspace{-0.2cm}
\noindent
\subsection*{\label{n28} 28}
In the Korteweg archives at University Library of University of Amsterdam.

\vspace{-0.2cm}
\noindent
\subsection*{\label{n29} 29}
In the De Vries archive maintained by the grandsons in Leiden.

\vspace{-0.2cm}
\noindent
\subsection*{\label{n30} 30}
The symbol ``${\sf F}$'' in the footnote, followed by ``\emph{Essai \dots}'', refers to the expression on line 12 from above. The translator of the present paper, Dr Behnam Farid, suggested that the expression concerns the \emph{total} derivative of $v\, [\equiv v(x,t)]$ with respect to $t$, where $x \equiv x(t)$; De Vries uses the symbols $d v/d t$ and $d v/d x$, rather than $\partial v/\partial t$ and $\partial v/\partial x$ respectively, for partial derivatives. The passage on which page 37 of De Vries' excerpt, shown in Fig.~\ref{f3}, has bearing, is directly taken from Saint-Venant's paper, page 1105. De Vries' exclamation mark in the footnote seems to indicate that he may have considered the derivation as significant.

\vspace{-0.2cm}
\noindent
\subsection*{\label{n31} 31}
De Jager, \emph{op. cit.}, 2006, 21.

\vspace{-0.2cm}
\noindent
\subsection*{\label{n32} 32}
I have not been able to find anything about this in the Korteweg archives, however as in the case of Brouwer, a great deal is as yet to be dug up.

}


\bibliographystyle{apsrev}


\end{document}